# The role of hydrodynamics for the spatial distribution of high-temperature hydrothermal vent-endemic fauna in the deep ocean environment

Zhiguo He[a,c,*,†], Yingzhong Lou[a,d,†], Haoyang Zhang[a], Xiqiu Han[a,b], Thomas Pähtz[a,*], Pengcheng Jiao[a,c], Peng Hu[a], Yadong Zhou[b], Yejian Wang[b], Zhongyan Qiu[b]

## Abstract

Active hydrothermal vents provide the surrounding submarine environment with substantial amounts of matter and energy, thus serving as important habitats for diverse megabenthic communities in the deep ocean and constituting a unique, highly productive chemosynthetic ecosystem on Earth. Vent-endemic biological communities gather near the venting site and are usually not found beyond a distance of the order of 100 m from the vent. This is surprising because one would actually expect matter ejected from high-temperature vents, which generate highly turbulent buoyancy plumes, to be suspended and carried far away by the plume flows and deep-sea currents. Here, we study this problem from a fluid dynamics perspective by simulating the vent hydrodynamics using a numerical model that couples the plume flow with induced matter and energy transport. We find that both low- and high-temperature vents deposit most vent matter relatively close to the plume. In particular, the tendency of turbulent buoyancy plumes to carry matter far away is strongly counteracted by generated entrainment flows back into the plume stem. The deposition ranges of organic and inorganic hydrothermal particles obtained from the simulations for various natural high-temperature vents are consistent with the observed maximum spatial extent of biological communities, evidencing that plume hydrodynamics exercises strong control over the spatial distribution of vent-endemic fauna. While other factors affecting the spatial distribution of vent-endemic fauna, such as geology and geochemistry, are site-specific, the main physical features of plume hydrodynamics unraveled in this study are largely site-unspecific and therefore universal across vent sites on Earth.

[a] Ocean College & Engineering Research Center of Oceanic Sensing Technology and Equipment of Ministry of Education, Zhejiang University, Zhoushan 316021, China
[b] Key Laboratory of Submarine Geosciences & Second Institute of Oceanography, Ministry of Natural Resources, Hangzhou 310012, China
[c] Hainan Institution, Zhejiang University, Sanya 572000, China
[d] Department of Civil and Environmental Engineering, University of Washington, Seattle, WA 98195, USA
[*] Corresponding authors. Email: hezhiguo@zju.edu.cn; 0012136@zju.edu.cn
[†] These authors contributed equally to this work.

## Graphical Abstracts

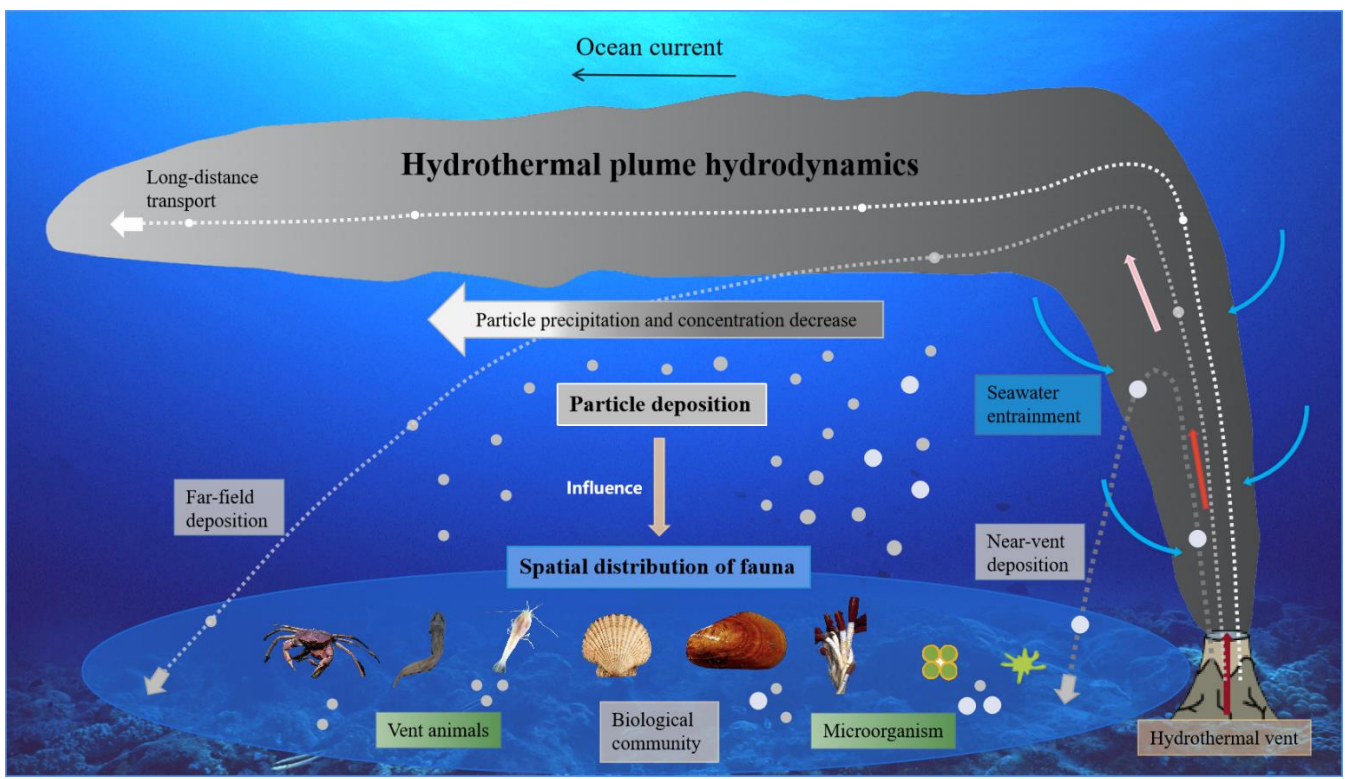

## Highlights:

1. We simulate the vent hydrodynamics and matter and energy transport through a coupled CFD and discrete phase model.
2. We characterize the hydrodynamic mechanisms and draw connections to vent-endemic species.
3. Hydrothermal particle range is consistent with the observed maximum spatial extent of endemic communities.
4. Hydrothermal plume hydrodynamics are site-unspecific and strongly influence fauna distribution.

## 1 Introduction

Since the discovery of hydrothermal vents and lush biological communities that live around active vent fields, suspected as the origin of life on Earth (Martin et al., 2008), vent (micro)organisms have been of great interest because they thrived in such high abundances despite low amounts of photosynthetically-derived food. (Tansey and Brock, 1972; Corliss et al., 1979; Cary et al., 1998). A summary of trophic interactions is shown in Fig. 1 (Galkin, 1997; Raguenes et al.,1997; Gamo et al., 2001; Emerson and Moyer, 2002; Inagaki et al., 2003; Tsurumi et al., 2003; Lopez-Gonzalez et al., 2005; Sancho et al., 2005; Voight, 2005; Duperron et al., 2006; Bergquist et al., 2007; Goffredi et al., 2008; Wang et al., 2021a).

The vent fields are an important medium for the exchange of heat and chemicals between the oceanic lithosphere and the overlying ocean; they are widespread not only along mid-ocean ridges but also in other geological settings, such as volcanic arcs, back-arc basins, and intraplate volcanoes (Fig. 1a) (Mullineaux, 2018; Beaulieu and Szafranski, 2020). A typical vent field contains a few high-temperature vents and numerous adjacent low-temperature vents (Mittelstaedt et al., 2012). Hydrothermal fluids ejected from high-temperature vents and rising up as a turbulent buoyancy plume rapidly mix with seawater, undergoing complex biotic and abiotic processes, such as complexation of dissolved metals (Bennett et al., 2008; Sander and Koschinsky, 2011) and cellular uptake of iron (Li et al., 2014), thus forming hydrothermal plumes containing various organic and inorganic metal-rich particles and their aggregations (Lilley et al., 1995; Baumberger et al., 2020; Tao et al., 2011).

Hydrothermal plumes can serve as the sources of both energy and matter not only for microorganisms (Lesniewski et al., 2012; Anantharaman et al., 2013) but also for diverse megabenthic communities at vent ecosystems (Grassle, 1985; Kelley et al., 2005; Fisher et al., 2007; Sen et al., 2013; Nakamura and Takai, 2014; Lee et al., 2015; Thomas et al., 2018; Watanabe et al, 2019), in spite of a high dilution of hydrothermal materials in the plume (approximately 1:10000) (Reed et al., 2015). In fact, the mixing of high-temperature plumes with cold seawater may occur in various ways and on different scales, resulting in diverse habitats (Dick, 2019) and thus greatly affecting existing organisms and their metabolic activities (Reysenbach et al., 2000; Orcutt et al., 2011). For example, temperature is shown to play a crucial role in determining the distribution of different organisms (Prieur et al., 1995; Reysenbach and Shock., 2002). While the upper temperature limit for biological growth, as observed in deep-sea hydrothermal Archaea, is about 122 °C (Takai et al.,2008), the majority of vent animals, such as tube worms, shrimp, crabs, bivalves, and snails, are unable to tolerate sustained temperatures above 55 °C (Girguis and Lee, 2006); they consequently inhabit regions with lower temperatures such as the periphery of high-temperature zones or around low-temperature vents. It should be noted that low-temperature vents are not associated with violent hydrodynamic processes due to their much smaller buoyancy and momentum. Therefore, released nutrients are expected to remain near the seabed surface adjacent to the vent.

It has been observed that vent-endemic animals always live within a certain distance (<~100 m) from hydrothermal vents and vent communities are characterized by a complex trophic and spatial structure (Fig. 1) (Galkin, 2016). This is surprising because one would actually expect matter ejected from a high-temperature vent, which serves as nutrients for vent-endemic fauna, to be suspended and carried far away by the vent plume and deep-sea currents. Several hypotheses have been proposed to resolve this conundrum (Renninger et al., 1995; Galkin, 1997; Kim and Hammerstrom, 2012). They usually invoke some kind of biochemical environmental characteristic (Johnson et al., 1986; Shank, 1998; Luther et al., 2001), such as the concentration of chemicals, and chemical speciation of the material emitted by the plume (e.g., oxygen, iron and sulphur speciation), which can strongly vary across vent sites (Luther et al., 2001). However, a more universal explanation would be one that is primarily based on plume hydrodynamics, since plumes across the globe can be roughly considered as self-similar, described by the source buoyancy flux and background stratification (Morton et al., 1956). For example, all hydrothermal plumes consist of large-scale vortices, generated by the shear between the inward entrainment flow around the stem of hydrothermal plumes and the outward transport of the plume's neutral buoyancy layer. It has been suggested that the trapping of vent larvae by such vortices is chiefly responsible for the spatial distribution of some biological communities (Mullineaux and France, 1995). If vent larvae can be trapped by plume vorticities, it seems reasonable that this can also be true for matter ejected from the vent, despite lacking the behavioral and physiological characteristics of the larvae. Even if not trapped, the entrainment flow may prevent ejected vent matter from travelling too far away, particularly when the background stratification is relatively strong (Lou et al., 2020). In fact, there are observations suggesting that the range of vents' spatial influence could be ~100 m (Thornton et al., 2016). Therefore, it should be reasonable to base estimations of the spatial extent of vent-endemic fauna on this scale of influence. Although experiments and conceptual models of hydrothermal vent plume-induced particle transport and deposition have been conducted (Sparks et al., 1991; Veitch and Woods, 2000; Zarrebini and Cardoso, 2000) and proposed (German and Sparks, 1993; Ernst et al, 1996; Dissanayake, 2014; Chan and Lee, 2016), the flow-driven trajectories of hydrothermal vent matter have only been investigated recently (Lou et al., 2020), and their effects on the spatial distribution of biological communities remain poorly understood.

Here, using simulations with a coupled computational fluid dynamics (CFD) and discrete phase model, we simulate the dynamics of hydrothermal plumes in a stratified ocean environment, summarize the transport patterns of particles of different particle sizes ($d_p$) in the plumes, and then characterize the hydrodynamic mechanisms that control the deposition range and spatial distribution of hydrothermal vent matter ejected from low- and high-temperature vents and draw connections to the spatial distribution of different vent-endemic species from various vent sites. Note that the simulation results depend predominantly on the settling velocity of the particles (Lou et al., 2020), which is a function of $d_p$ and the particle density $\rho_p$. Since the relevant range of $d_p$ covers about three orders of magnitude for both organic and inorganic particles, whereas $\rho_p$ varies only about by a factor of 4 between organic and inorganic particles (Alldredge and Crocker, 1995; Breier et al., 2012; Haynes et al., 2016), and since the effect of $d_p$ on the settling velocity is much larger than that of $\rho_p$ (Lou et al., 2020), we choose to show below only plots for representative particle species across the range of relevant particle sizes. In order words, results shown for inorganic particles apply to organic particles as well and vice versa.

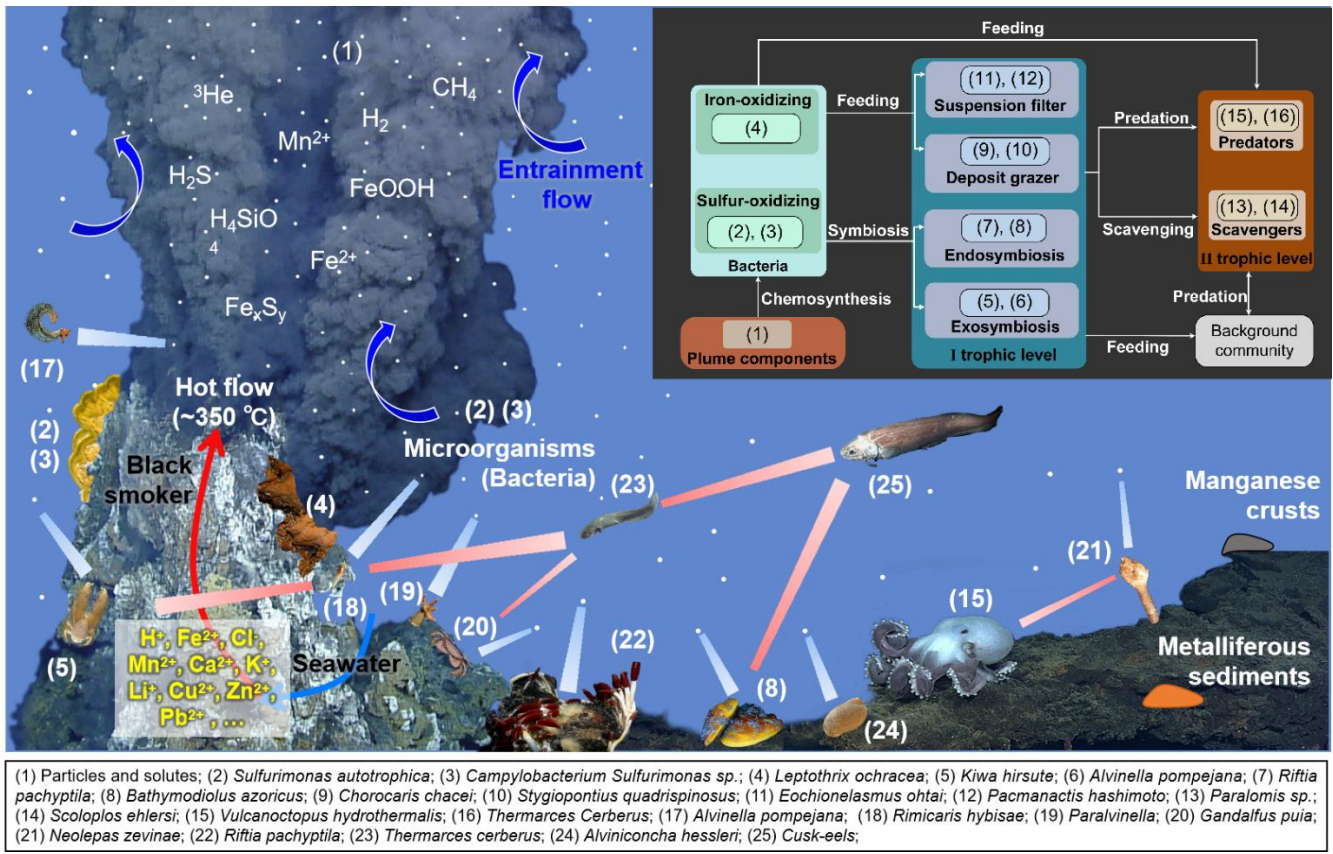

**Fig. 1**. **Schematic of the food web in submarine hydrothermal ecosystems.** Note that the blue bars indicate the utility of hydrothermal organic particles and solutes for animals via microorganisms, while the red bars represent consumption. The size and color gradient are purely for visual display purposes.

## 2 Materials and Methods

### 2.1 Hydrodynamic model

Simulations with the hydrodynamic model of hydrothermal plumes developed in this study are carried out using the finite-volume CFD software package ANSYS Fluent. It solves the Reynolds-averaged mass, momentum, and energy conservation equations. In Cartesian coordinates, they can be expressed as (Einsteinian summation):

$$\frac{\partial \rho}{\partial t} + \frac{\partial \rho u_i}{\partial x_i} = 0 \tag{1}$$

$$\frac{\partial \rho u_i}{\partial t} + \frac{\partial \rho u_i u_j}{\partial x_j} = -\frac{\partial p}{\partial x_i} + \frac{\partial \tau_{ij}}{\partial x_j} + \frac{\partial}{\partial x_j}\left(-\rho \overline{u'_i u'_j}\right) + \rho g_i \tag{2}$$

$$\frac{\partial \rho E_m}{\partial t} + \frac{\partial}{\partial x_j}\left[u_j\left(\rho E_m + p\right)\right] = \frac{\partial}{\partial x_j}\left(\kappa_{eff} \frac{\partial T}{\partial x_j}\right) + \frac{\partial}{\partial x_j}\left(u_i \frac{\mu_{eff}}{\mu} \tau_{ij}\right) \tag{3}$$

where $t$ is the time, $x_i$ is the spatial coordinates, $\rho$ is the fluid density, $u_i$ is the average velocity, $p$ is the pressure, $\tau_{ij} = \mu\left(\frac{\partial u_i}{\partial x_j} + \frac{\partial u_j}{\partial x_i} - \frac{2}{3}\delta_{ij}\frac{\partial u_l}{\partial x_l}\right)$ is the strain rate tensor, $\mu$ is the viscosity, $u'_i$ is the fluctuating velocity, $g_i$ is the body force, $E_m$ is the total energy per unit mass of fluid, $\kappa_{eff}$ is the effective thermal conductivity, $T$ is the thermodynamic temperature, $\mu_{eff}$ is the effective viscosity (i.e. the sum of molecular viscosity and turbulent viscosity), and $\delta_{ij}$ is the Kronecker delta. $-\rho \overline{u'_i u'_j}$ are Reynolds stresses, which can be modeled in Cartesian tensor form using the Boussinesq hypothesis as:

$$-\rho \overline{u'_i u'_j} = \mu_t\left(\frac{\partial u_i}{\partial x_j} + \frac{\partial u_j}{\partial x_i}\right) - \frac{2}{3}\delta_{ij}\left(\rho k + \mu_t \frac{\partial u_k}{\partial x_k}\right) \tag{4}$$

where $\mu_t = \rho C_\mu k^2 / \varepsilon$ is the turbulent vortex viscosity, $k$ is the turbulent kinetic energy, and $\varepsilon$ is the turbulent dissipation rate. $C_\mu = \left(A_0 + A_S \dfrac{kU^*}{\varepsilon}\right)^{-1}$ is a coefficient as a function of the mean strain and rotation rates and the turbulence fields, where $U^* = \sqrt{S_{ij}S_{ij} + \Omega_{ij}\Omega_{ij}}$, $\Omega_{ij} = \dfrac{1}{2}\left(\dfrac{\partial u_i}{\partial x_j} - \dfrac{\partial u_j}{\partial x_i}\right)$ is the rate-of-rotation tensor, $A_0 = 4.04$, $A_S = \sqrt{6}\cos(\phi)$, $\phi = \dfrac{1}{3}\cos^{-1}(\sqrt{6}\omega)$, $\omega = \dfrac{S_{ij}S_{jl}S_{li}}{(S/\sqrt{2})^3}$, $S = \sqrt{2S_{ij}S_{ij}}$, and $S_{ij} = \dfrac{1}{2}\left(\dfrac{\partial u_i}{\partial x_j} + \dfrac{\partial u_j}{\partial x_i}\right)$ is the rate-of-strain tensor.

A $k$-$\varepsilon$ model (Shih et al., 1995) is employed in this study. The transport equations for $k$ and $\varepsilon$ in the model are:

$$\frac{\partial}{\partial t}(\rho k) + \frac{\partial}{\partial x_j}(\rho k u_j) = \frac{\partial}{\partial x_j}\left[\left(\mu + \frac{\mu_t}{\sigma_k}\right)\frac{\partial k}{\partial x_j}\right] + G_k + G_b - \rho\varepsilon \quad (5)$$

$$\frac{\partial \rho\varepsilon}{\partial t} + \frac{\partial \rho\varepsilon u_j}{\partial x_j} = \frac{\partial}{\partial x_j}\left[\left(\mu + \frac{\mu_t}{\sigma_\varepsilon}\right)\frac{\partial \varepsilon}{\partial x_j}\right] + \rho C_1 S\varepsilon - \rho C_2 \frac{\varepsilon^2}{k + \sqrt{\nu\varepsilon}} + C_{1\varepsilon}\frac{\varepsilon}{k}C_{3\varepsilon}G_b \quad (6)$$

where $G_k = -\rho\overline{u'_i u'_j}\dfrac{\partial u_j}{\partial x_i} = \mu_t S^2$ represents the generation rate of turbulent kinetic energy due to the mean velocity gradients, with $S = \sqrt{2S_{ij}S_{ij}}$ the modulus of the mean rate-of-strain tensor and $S_{ij} \equiv \dfrac{1}{2}\left(\dfrac{\partial u_i}{\partial x_j} + \dfrac{\partial u_j}{\partial x_i}\right)$ the mean strain rate. $G_b = \beta g_i \dfrac{\mu_t}{Pr_t}\dfrac{\partial T}{\partial x_i}$ is the generation rate of turbulent kinetic energy due to buoyancy, with $Pr_t = 0.85$ the turbulent Prandtl number for energy and $\beta = -\dfrac{1}{\rho}\left(\dfrac{\partial \rho}{\partial T}\right)_P$ the coefficient of thermal expansion. $C_{3\varepsilon} = \tanh\left(\left|\dfrac{w}{v}\right|\right)$ determines the degree to which $\varepsilon$ is affected by the buoyancy, and $w$, $v$ are

the components of the flow velocity parallel and perpendicular to the gravitational vector, respectively. $\sigma_k = 1.0$ and $\sigma_\varepsilon = 1.2$ are the turbulent Prandtl numbers for $k$ and $\varepsilon$, respectively. $C_1 = \max\left[0.43, \frac{\eta}{\eta+5}\right]$, $C_2 = 1.9$, $C_{1\varepsilon} = 1.44$ are model constants with $\eta = S\frac{k}{\varepsilon}$.

Furthermore, the effective thermal conductivity $\kappa_{eff}$ in the model is estimated to be

$$\kappa_{eff} = \kappa + \frac{C_p \mu_t}{Pr_t} \tag{7}$$

where $\kappa$ is the thermal conductivity of the effluent and $C_p$ is its specific heat capacity. Note that the specific heat $C_p$ and thermal conductivity $\kappa$ are taken as functions of temperature, salinity, and pressure, and their parameterization will be discussed below. At the same time, a scalar quantity $\phi$ is introduced to describe the concentration distribution of the components in the plume during its evolution, and the transport and diffusion of the plume are estimated using the following equations:

$$\frac{\partial \rho \phi}{\partial t} + \frac{\partial}{\partial x_i}\left(\rho u_i \phi - \Gamma \frac{\partial \phi}{\partial x_i}\right) = S_\phi \tag{8}$$

where $\Gamma$ is the diffusion coefficient of the substance (will be discussed below) and $S_\phi$ is the source term of the substance, respectively.

## 2.2 Model configuration

This study simulates the flow field of a hydrothermal plume with and without ocean currents. The computational domain of the model for the cases without ocean currents is axisymmetric, where the centerline of the hydrothermal vent is taken as the symmetry axis. The domain, whose horizontal span and depth are 1000 m and 600 m respectively, is discretized by triangular cells. Suzuki et al. (2005) showed that grid sizes smaller than $D_0/10$ correctly reproduce turbulent mixing near the vent, where $D_0$ is the vent diameter. Therefore, our vent radius is evenly divided into 10 segments, in accordance with previous studies (Suzuki et al., 2005; Suzuki & Koyaguchi, 2010; Lou et al., 2020). The top of the domain is set as a pressure-outlet boundary with a constant temperature and pressure, and the outer vertical boundary is considered as a symmetric boundary.

The computational domain of the model for the cases with ocean currents is a cuboid whose XZ plane is taken as the symmetry plane. The domain, with a length 1500 m, width 1000 m, and height 600 m, is fully discretized by tetrahedral cells, and the center of the vent orifice is located at (300,0,0). The top of the domain is set as a pressure-outlet boundary with a constant temperature and pressure, and the YZ plane is the inlet of the ocean currents.

Since the variations of both salinity and pressure within the main flow domain are usually small, the temperature could be regarded as the only controlling factor of plume properties in the present model. Specifically, fluid density ($\rho$), molecular viscosity ($\mu$), specific heat ($C_p$) and thermal conductivity ($\kappa$) are taken as functions of temperature ($T$) at fixed salinity and pressure following previous studies (Sun et al., 2008; Jiang and Breier, 2014).

## 2.3 Initial and boundary conditions

The initial temperature field is set to decrease linearly with the depth, thus leading to an almost constant background buoyancy frequency

$$N = \sqrt{-\frac{g}{\rho_{bottom}}\frac{\partial \rho}{\partial Z}} \qquad (9)$$

where $\rho_{bottom}$ is fluid density at the bottom of the domain and $g=9.81$ m/s² is the gravitational acceleration. Moreover, the source buoyancy flux of the hydrothermal plume is

$$B_{exit} = g \frac{\rho_{bottom} - \rho_{exit}}{\rho_{bottom}} Q_{exit} \tag{10}$$

where $\rho_{bottom}$ and $\rho_{exit}$ are the fluid densities at the bottom of the ocean and at the vent orifice, respectively, and $Q_{exit}$ is the source volume flux.

It is noteworthy that the effluent flow from a hydrothermal vent is influenced by two important factors: buoyancy and initial momentum flux, particularly in the near-field region. Consequently, the entrainment process around the plume stem may differ from that of a pure plume and exhibits similarities to a forced plume (or turbulent jet). In the present model, the possible particle concentration is assumed to be low and thus chemical reactions and diffusion of solutes are omitted since the simulation time scale is relatively small and advection is dominant.

## 2.4 Particle transport model

After obtaining the hydrothermal plume flow field through the computational fluid dynamics model, a discrete phase model is coupled with it to predict the characteristic trajectory of transported hydrothermal source particles in the marine stratified environment. If the interaction between hydrothermal particles is assumed to be negligible, the equations of motion of spherical particles in cylindrical coordinates can be written as:

$$\frac{d\vec{u}_p}{dt} = \frac{\vec{u} - \vec{u}_p}{\tau_r} + \frac{\rho_p - \rho}{\rho_p} \vec{g} + \vec{F}_a \tag{11}$$

where $\vec{u}_p$ is the particle velocity, $\vec{u}$ is the velocity of the fluid, $\vec{g}$ is the gravitational acceleration, and $\vec{F}_a$ is the additional forces per unit mass (and will be described below).

The particle relaxation time $\tau_r$ is defined as (Gosman and Ioannides, 1983)

$$\tau_r = \frac{\rho_p d_p^2}{18\mu} \frac{24}{C_d Re} \tag{12}$$

where $Re$ is the relative Reynolds number and $C_d$ is the drag coefficient. The latter two are defined as (Yick et al., 2009):

$$Re = \frac{\rho d_p}{\mu} |\vec{u}_p - \vec{u}| \tag{13}$$

$$C_d = \left(\frac{24}{Re} + \frac{6}{1+\sqrt{Re}} + 0.4\right)\left(1 + \beta Ri_\mu^{1/2}\right) \tag{14}$$

where $\beta = 1.9$ is an empirical coefficient, and $Ri_\mu$ is the viscous Richardson number, defined as:

$$Ri_\mu = \frac{\rho d_p^3}{8\mu |\vec{u}_p - \vec{u}|} \left|\frac{g}{\rho_{bottom}} \frac{\partial \rho}{\partial z}\right| \tag{15}$$

In the model, the virtual mass force, pressure gradient force, and Saffman's lift force are considered as the main additional external forces, while other potentially important external forces are neglected. Specifically, since this model only considers solid particles moving in a liquid environment, it is reasonable to neglect the thermal force and Brownian motion. Furthermore, for conditions with a fluid-to-particle density ratio in the range of $0.1 < \rho/\rho_p < 1$, it has been shown (Michaelides, 1997) that ignoring the Basset history force usually does not result in significant errors when the background flow field is stable and the spatial variation rate of the fluid velocity is low. In summary, the external force per unit mass employed in this model can be expressed as (Saffman, 1965; Li and Ahmadi, 1992):

$$\vec{F}_a = \vec{F}_{vm} + \vec{F}_p + \vec{F}_s \tag{16}$$

where $\vec{F}_{vm}$ is the virtual mass force, $\vec{F}_p$ is the pressure gradient force, and $\vec{F}_s$ is the Saffman's lift force. They are defined as:

$$\vec{F}_{vm} = C_{vm} \frac{\rho}{\rho_p} \left( \vec{u}_p \nabla \vec{u} - \frac{d\vec{u}_p}{dt} \right) \qquad (17)$$

$$\vec{F}_p = \frac{\rho}{\rho_p} \vec{u}_p \nabla \vec{u} \qquad (18)$$

$$\vec{F}_s = \frac{2K\sqrt{\mu\rho}d_{ij}}{\rho_p d_p \left( d_{lk} d_{kl} \right)^{1/4}} \left( \vec{u} - \vec{u}_p \right) \qquad (19)$$

In Eq. (17), $C_{vm} = 0.5$ is the shape parameter; in Eq. (19), $K = 2.594$ is a constant and $d_{ij}$ is the deformation tensor. It should be pointed out that, since the Saffman's lift force is only important for small particle Reynolds number, it is only applied to submicron-sized particles in this model.

# 3 Results and Discussion

We use the coupled CFD and discrete phase model presented above to simulate vent hydrodynamics and matter and energy transport.

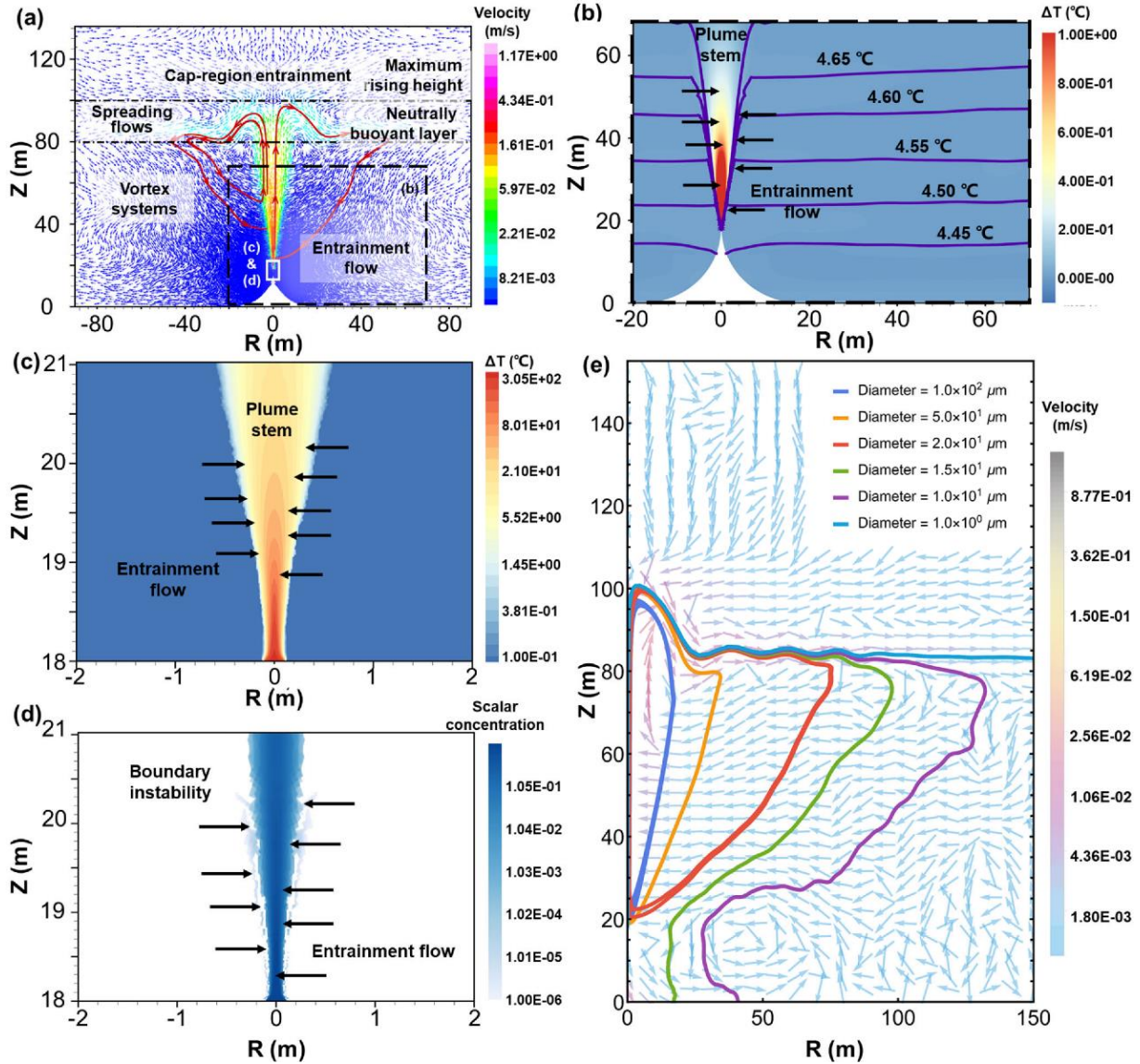

**Fig. 2. Hydrodynamics of hydrothermal plumes in the calm deep ocean.** (**a**) Velocity field and typical streamlines of a hydrothermal plume, where color (not length) indicates the velocity magnitude. (**b**) Temperature anomaly ΔT=T(R,Z)-$T_0$(Z) in the near vent region, where $T_0$(Z) is the background temperature and the purple contours are local temperature. (**c**) The temperature anomaly and (**d**) scalar concentration

in the region above the vent. (**e**) Trajectories of goethite particles of different diameters, where color indicates the velocity magnitude. Note that this pattern is representative of all kinds of organic and inorganic particles. Note that the color of the background and particles is distinguished by transparency.

## 3.1 General hydrodynamic features of a hydrothermal vent

A high-temperature hydrothermal vent is defined as a vent that emits high-temperature fluid with an initial velocity into the denser low-temperature sea water, forming a rising plume (Fig. 2a). During the initial phase of the rise, the plume accelerates because the buoyancy force more than compensates the opposing gravity force. However, while the plume rises, more and more cold ambient sea water is entrained into the plume stem due to compensation flow, causing the rising plume to cool down and decelerate following its initial acceleration. The more the plume cools down, the denser it becomes and the weaker becomes the buoyancy force. At some point, the net force becomes negative, forcing the plume water to bounce back and spread sideways. The inward entrainment flow combined with this plume's spreading flow promotes the formation and growth of plume vortices. All these processes also occur for low-temperature vents, however to a much smaller degree due to much smaller buoyancy and initial momentum (e.g., a buoyancy plume is hardly detectable).

## 3.2 Solutes distribution and particle deposition from high-temperature vents

The simulations show that the general circulation system tends to concentrate solutes and heat energy below the neutrally buoyant layer in the immediate vent vicinity (Fig. 2b-d). This holds true even in the presence of background flows within a few centimeters per second (Fig. 3a), associated with deep ocean currents. Since solutes are the primary nutrient source for chemoautotrophic microorganisms (e.g., iron-oxidizing and sulfur-oxidizing bacteria), this finding offers a potential explanation for the previously observed spatial limitation of some microorganisms to the very-near vent region in natural vent systems (Gundersen et al., 1992; Galkin, 2016).

We distinguish three regimes of particle transport and deposition caused by the general circulation pattern (Fig. 2e). They depend on the settling velocity $\omega_s$ and thus on the size and density of a given particle emitted from the vent. First, the ascending motion of the largest particles emitted from the vent (largest $\omega_s$) is turned into a descending motion before they reach the spreading flow. They are then reentering the high-velocity plume stem, preventing them from depositing (i.e., they remain in suspension; note that the impact of background flow will be discussed below). Second, intermediately-sized particles emitted from the vent (intermediate $\omega_s$) reach the spreading flow and are then deflected towards the plume by the entrainment flux during their settling process, causing large deposition fluxes in the near-vent region (near-vent deposition). Third, small particles emitted from the vent (small $\omega_s$) remain suspended in the spreading flow, are transported away from the vent to very large distances (long-distance transport), and therefore are highly diluted and do not serve as food for animals living at the sea floor near the vent. In particular, many of them will be decomposed long before deposition due to the long traveling times. Crucially, the fact that the transition between near-vent deposition and long-distance transport is generally abrupt motivates using the farthest distance travelled by particles in the near-vent deposition regime as a measure for the deposition range and therefore as a predictor of the maximum spatial extent of biological communities.

Background flows can substantially extend the deposition range of hydrothermal particles and change their deposition patterns (Fig. 3b, note that the particle size selection in Figs. 2e and 3b is not exactly the same, mainly to reflect representative trajectories). In particular, the largest particles may be able to leave the plume stem and deposit downstream, but not upstream, in the direction of the current in the near-vent region, thus potentially causing an inhomogeneous distribution of biological communities in the circumferential direction (Thornton et al., 2016; Gerdes et al., 2019b). Intermediately-sized particles will still reach the spreading flow, but then, instead of being deflected towards the plume, they will be transported away from the plume for sufficiently strong background flows, resulting in far-field deposition. However, as discussed above, far-field-deposited organic particles are usually not an important nutrient source for vent-endemic biological communities. Small particles can still be suspended in the spreading flow and be transported to very large distances when background flows are present.

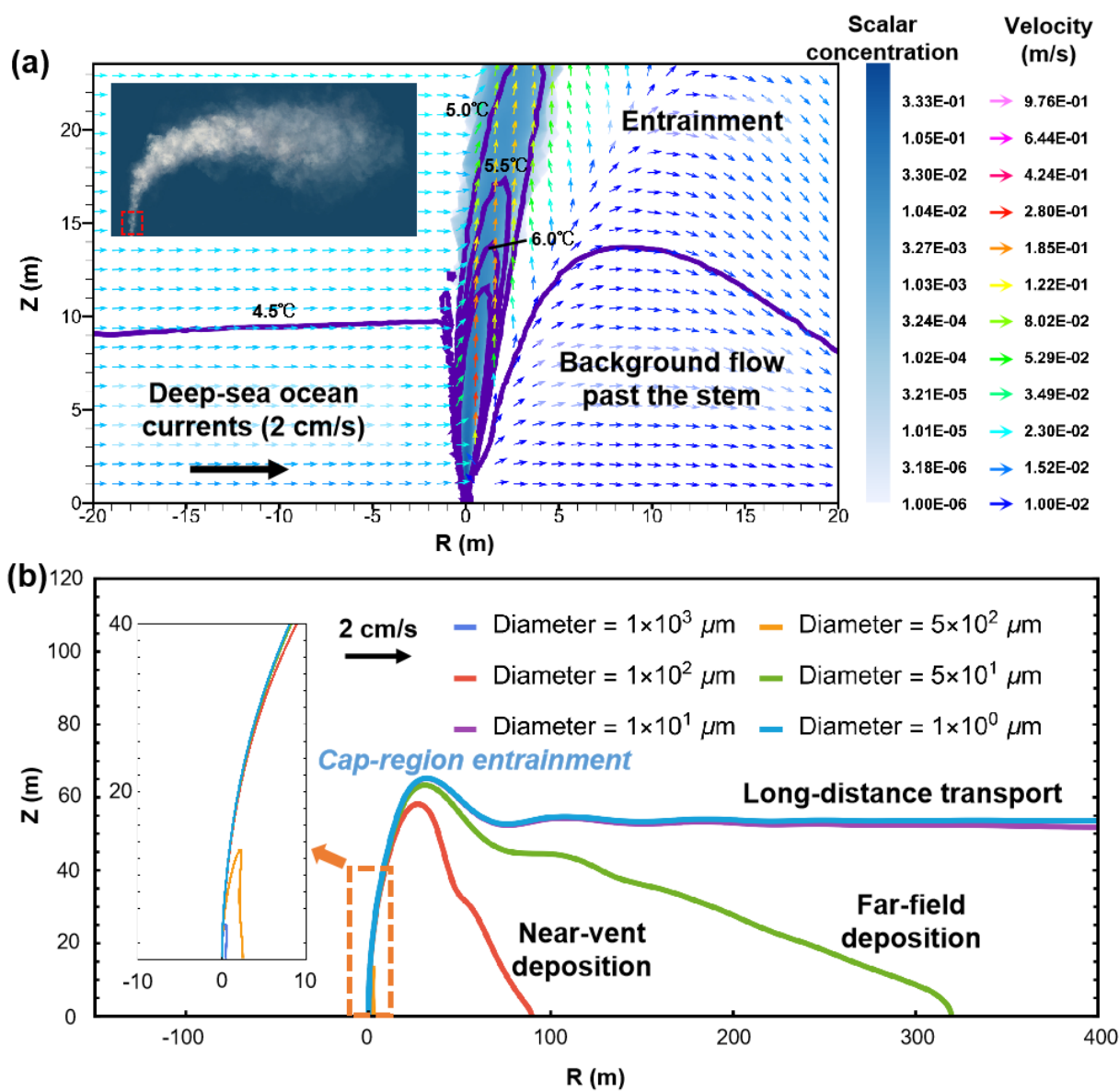

**Fig. 3. Hydrodynamics of hydrothermal plumes in the deep ocean with currents.** (**a**) Modeled velocity

field and solute concentration field in the near vent region of the hydrothermal plume under the influence of 2 cm/s ocean currents. The purple contour lines indicate temperature; the inset shows the lateral transport of the plume by the current. Note that here we present a representative scenario with background flows; a more detailed discussion of the hydrodynamics in the presence of background flows can be found in Lou et al. (2020). (**b**) The trajectories of particles with a density of 5020 kg/m$^3$; the inset shows the detailed trajectories in the near-vent region marked by the orange box.

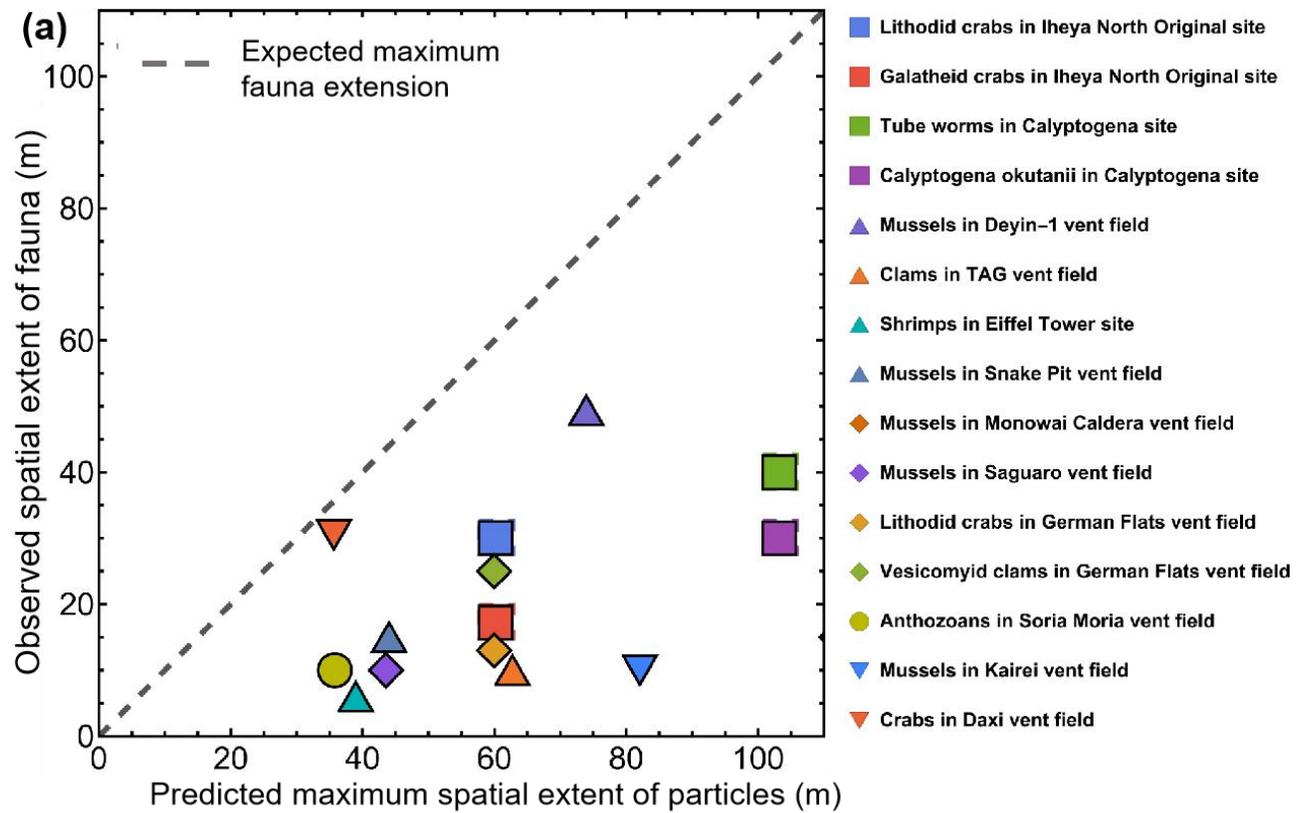

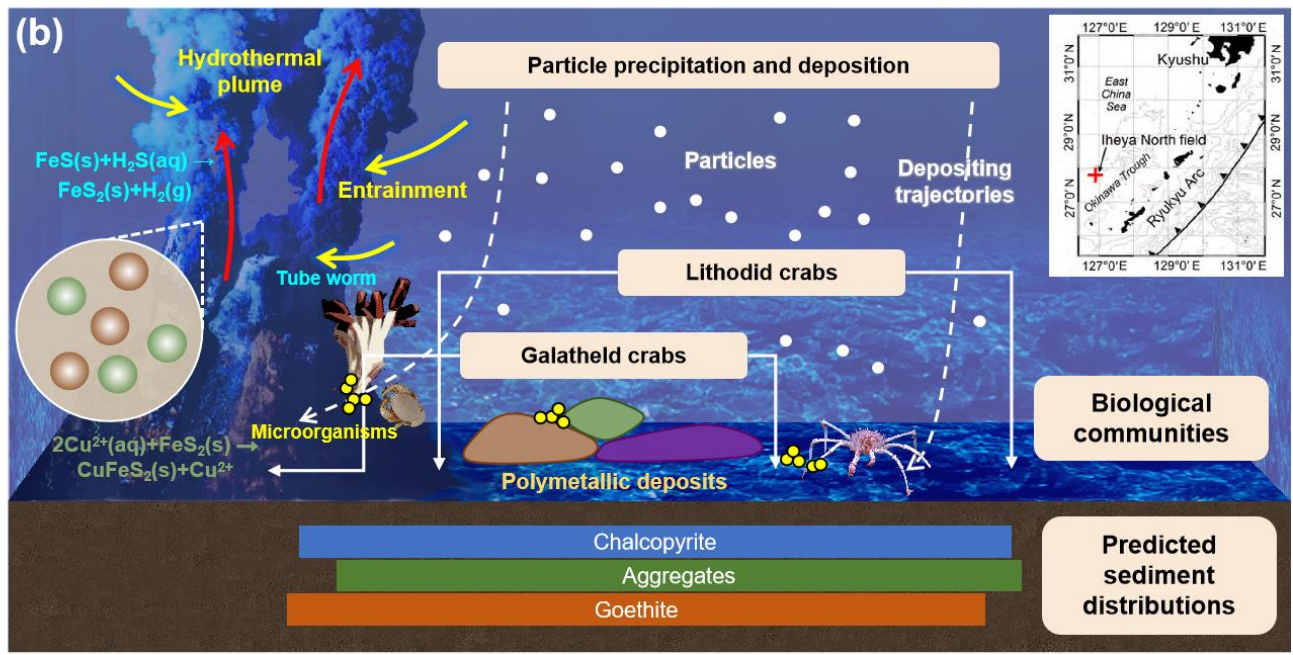

**Fig. 4. Fauna distribution and schematic of energy transport through hydrothermal particles.** (a) Observed spatial extent of vent fauna versus predicted maximum spatial extent of particles in natural vent sites. The fact that all data are below (but not much below) the dashed line supports the hypothesis

that vent hydrodynamics control the spatial distribution of biological communities. Data of hydrodynamic conditions and animal distributions are adopted from World Ocean Atlas 2018 (Boyer, 2018) and field observations (Van Dover, 1995; Ohta and Kim, 2001; Stecher et al., 2002; Turnipseed et al., 2004; Hey et al., 2006; Cuvelier et al., 2009; German et al., 2010; Leybourne et al., 2012; Doostmohammadi et al., 2014; Thornton et al., 2016; Dahle et al., 2018; Gerdes et al., 2019a; Adam et al., 2020; Wang et al., 2021b). (**b**) Distributions of deposited particles and representative vent fauna in NBC (North Big Chimney) region. Animal distributions are adopted from the reconstructions (Thornton et al., 2016). The regional map is adapted from Thornton et al. (2016).

Deep ocean currents exhibit variability at a characteristic time of a few hours, which is much shorter than the typical time needed for vent-endemic biological communities to adapt to the vent environment (Jiang, 2017). We assume that the transport of nutrients is predominantly controlled by the long-term-average deep ocean current, and, as a first approximation, we assume that this average current velocity is sufficiently weak, so that it is justified to use a model with no background flow. When applied to natural vent sites, this simplified model predicts a maximum spatial extent of particle deposition that is of the same order of magnitude as observed extents of biological communities (Fig. 4a). Note that the observed extents referred to here are the maximum values among the reports regarding local biological communities. This quantitatively supports our main hypothesis that the spatial distribution of vent-endemic biological communities is predominantly controlled by plume hydrodynamics. Furthermore, to visualize the accumulation of hydrothermal particles such as chalcopyrite (5020 kg/m$^3$), goethite (4300 kg/m$^3$), and organic aggregates (1100 kg/m$^3$), we show their numerically calculated distributions exemplary for a hydrothermal vent in the North Big Chimney (NBC) region (Fig. 4b). Note that the slight difference between the deposition ranges of the three particle types can be attributed to the resolution of particle size in the present simulations, which is set at 1 μm. In other words, the model predicts almost the same deposition ranges for both organic food particles and mineral particles.

## 4 Conclusions

In this study, we have hypothesized a strong connection between hydrodynamic features of deep-sea hydrothermal plumes, the concentration of vent-produced hydrothermal solutes, near-vent deposition of particles, and the distribution of vent-endemic fauna. To test this hypothesis, we have carried out simulations with a model coupling plume hydrodynamics and particle transport. In the presence of a weak mean background flow, the maximum spatial extent of deposited particles predicted by the model, which corresponds to the maximum spatial extent of vent-endemic biological communities, is of the same order of magnitude but, importantly, remains above the observed extent of biological communities around high-temperature vents. Crucially, in contrast to other site-specific factors affecting the spatial distribution of vent-endemic fauna, such as geology and geochemistry, the main physical features of plume hydrodynamics are largely site-unspecific (i.e., general across vent sites) due to geometric similarity of existing vent sites on Earth. Interestingly, the model reveals that solutes and heat energy are always concentrated in the immediate vent vicinity (<~10 m), regardless of deep ocean currents, since they remain trapped in the plume stem due to entrainment flow at the stem's boundary. This has the implication that microorganisms that use solutes as their main food source, such as sulfur-oxidizing bacteria, can only thrive in the immediate vicinity of the plume stem. Future studies may be able to test this prediction, which would lend further credence to the working hypothesis of this paper.

## CRediT authorship contribution statement

**Zhiguo He:** Conceptualization, Methodology, Resources, Supervision, Writing - Review & Editing, Project administration, Funding acquisition. **Yingzhong Lou:** Methodology, Software, Validation, Formal analysis, Investigation, Writing - Original Draft. **Haoyang Zhang:** Methodology, Software,

Writing - Review & Editing. **Xiqiu Han:** Investigation, Resources. **Thomas Pähtz:** Formal analysis, Writing - Original Draft. **Pengcheng Jiao:** Writing - Review & Editing. **Peng Hu:** Formal analysis, Writing - Review & Editing. **Yadong Zhou, Yejian Wang and Zhongyan Qiu:** Writing - Review & Editing.

## Data and materials availability

The data that support the findings of this study are available from the corresponding author upon reasonable request.

## Declaration of competing interest

The authors declare that they have no known competing financial interests or personal relationships that could have appeared to influence the work reported in this paper.

## Acknowledgments

This work was financially supported by the National Key R&D Program of China (Grant No. 2021YFF0501302), National Natural Science Foundation of China (Grant Nos. 52171276, 91951201, 12272344); Key R&D Program of Zhejiang Province (2021C03180); Fundamental Research Funds for the Central Universities (Grant No. 2017XZZX001-02A). This work was supported by HPC Center OF ZJU (ZHOUSHAN CAMPUS). We thank the three anonymous reviewers for their valuable and insightful comments and suggestions that have helped us improve the quality of the manuscript.